# TIME SYNCHRONIZATION PROTOCOL FOR THE KLJN SECURE KEY EXCHANGE SCHEME

LASZLO B. KISH

*Department of Electrical and Computer Engineering, Texas A&M University*
*TAMUS 3128, College Station, TX 77843, USA*

*Laszlokish@tamu.edu*

The information theoretically secure Kirchhoff-law-Johnson-noise (KLJN) key exchange scheme, similarly to quantum key distribution (QKD), is also potentially vulnerable against clock attacks, where Eve takes over the control of clock synchronization in the channel. This short note aims to introduce a time synchronization protocol scheme for Alice and Bob, which is resistant against arbitrary time delay attacks, both symmetric and asymmetric ones. We propose and explore various ways of clock synchronization for the KLJN system and propose an ultimate protocol that preserves time and hardware integrity under arbitrary attacks.

## 1. Introduction: the KLJN scheme

### 1.1 Information-theoretic (unconditional) security and time synchronization

On May 4, 2022, President Joe Biden signed a National Security Memorandum [1] that promotes (among others) the research and development of quantum-resistant cryptography. Such a crypto requires information-theoretic (that is, unconditional) security [2,3] of the secure key exchange protocol. That means that the privacy is not based on a mathematically hard problem, where *the hardness of that problem is only an assumption* - without a proof - such as in the case of all the conditionally secure schemes widely used nowadays. An unconditionally secure scheme requires a hardware solution where the laws of physics guarantee the privacy against passive (listening) attacks of the eavesdropper (Eve).

The Kirchhoff-law-Johnson-noise (KLJN) key exchange system [4-61] offers unconditional security by utilizing classical statistical physics (and the Second Law of Thermodynamics), while Quantum Key Distribution (QKD) [62-99] uses quantum physics (and the Quantum No-Cloning Theorem) to reach this goal.

Unfortunately, both these protocols are potentially vulnerable to active attacks where Eve takes over the control of the clocks of the communicating parties, Alice and Bob (see [100-112] about QKD). In this paper we propose a general defense against arbitrary clock attacks in the KLJN system. First the KLJN protocol is outlined below.

*1.2 On the KLJN secure key exchange protocol*

The core of the KLJN scheme is shown in Figure 1. Alice and Bob have identical pairs of resistors, $R_L$ (low resistance) and $R_H$ (high resistance), respectively. The four resistors have their independent Johnson voltage noise generators $U_{LA}(t)$, $U_{HA}(t)$, $U_{LB}(t)$, $U_{HB}(t)$, respectively, which are bandlimited Gaussian white noises with identical bandwidth $B$. At the beginning of each *bit exchange period* (BEP), Alice and Bob randomly and independently chose one of the resistors and connect their chosen resistors to the cable (wire channel). The connected voltage generators yield the cable voltage $U_c(t)$ and current $I_c(t)$, and their respective power density spectra $S_u(f)$ and $S_i(f)$. To evaluate the bit situation, Alice, Bob and Eve are measuring the mean-square values of $U_c(t)$ and $I_c(t)$.

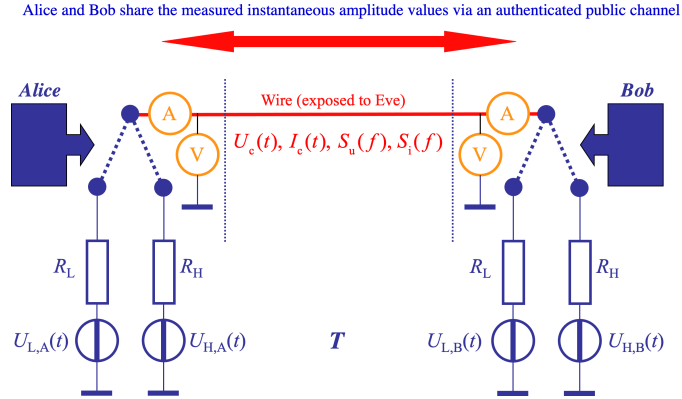

Figure 1. The core KLJN system with generic defense against active attacks. Alice and Bob have identical pairs of resistors. At the beginning of each bit *exchange period* (BEP), Alice and Bob randomly and independently choose one of the resistors $R_L$ (low resistance) or $R_H$ (high resistance) and connect their chosen resistors to the cable (wire channel). The four resistors have their independent Johnson voltage noise generators $U_{LA}(t)$, $U_{HA}(t)$, $U_{LB}(t)$, $U_{HB}(t)$, respectively, which are bandlimited white noises with identical bandwidth $B$. The connected voltage generators yield the cable voltage $U_c(t)$ and current $I_c(t)$, and their respective power density spectra $S_u(f)$ and $S_i(f)$, respectively. The measured current and voltage data are shared and compared/validated by Alice and Bob via authenticated communication.

Secure bit exchange takes place when the choice of the resistors is mixed: LH or HL because then the mean-square voltage and current values in the cable are identical. These two alternatives provide the secure bit values. For example, Alice and Bob agree that LH means 0 key bit value while HL interpreted as bit value 1. The security guaranteed by the fact that, in the mixed resistor case, Eve cannot distinguish between LH or HL. On the other hand, Alice and Bob know their own connected resistor values, thus they can also determine the choice of the other party.

Various attack types have been proposed against the KLJN scheme [4-10, 41-61], and all these attacks have one or more corresponding defense methods. The general defense method against active (invasive) attacks, i.e. attacks that inject or extract energy from the line and/or change the line, is based on the authenticated exchange of the current and voltage measurement data of Alice and Bob. Then, in the simplest case, they compare these data and evaluate their integrity. In the most advanced method, they run their own cable simulators with the actual parameters of the cable, and then, for example, check if the simulated current results are identical to the measured current data while the measured voltages are the input data of the simulators.

The same defense method is enhanced for the synchronization of the clocks, that is, to defend against the clock attacks, see in Section 3.3.

## 2. The problem of clock attacks

If Eve takes over the control of clocks in physical key exchanges schemes that means she can control time - a situation, which is equivalent becoming a "demigod" over the physical process. Such a situation opens the possibility of attacks. Therefore, in quantum key distribution (QKD) schemes the issue of clock synchronization is one of the highest importance [99-111]. This is complicated by the fact that QKD needs time synchronization with high accuracy, in the *picosecond* range.

No clock attack method has been published against the KLJN system yet. On the other hand, it is easy to see that synchronization is important. For example, if the clocks of Alice and Bob mismatched, the general defense method against active attacks mentioned above would almost continuously send alarm warnings even when there was no attack present.

Luckily, for the KLJN scheme to preserve its integrity and security, the time synchronization needs a time resolution that is much less demanding than for QKD. That follows from the autocorrelation function $R(\tau)$ of the white noise voltage $U(t)$ limited in frequency bandwidth $B$, where $\tau$ is the time shift variable:

$$R(\tau) \equiv \langle U(t)U(t+\tau) \rangle = BS_0(f)\frac{\sin(2\pi B\tau)}{2\pi B\tau} \quad . \tag{1}$$

Equation 1 is plotted in Figure 2. It is obvious that "virtually nothing" happens within a small fraction of the time interval of the inverse bandwidth $1/B$ thus the accuracy of the time synchronization must be in this range. That practically means a resolution in the order of the flying time of EM waves in the cable (information channel).

For the Reader unfamiliar with the KLJN system here we mention that, in typical practical situations the following approximate relations are feasible [20,49]:

$$1000\tau_f \approx \frac{100}{B} \approx \text{BEP},  \qquad (2)$$

where $\tau_f$ is the flying time of EM waves between Alice and Bob through the cable, so $\tau_f \approx \frac{0.1}{B}$. For example, in the case of a 2-kilometer range, the resolution of time synchronization needs to be in the order of 10 microsecond, which is a convenient requirement and about a million times longer than it is required for QKD. The only open question is how to achieve that.

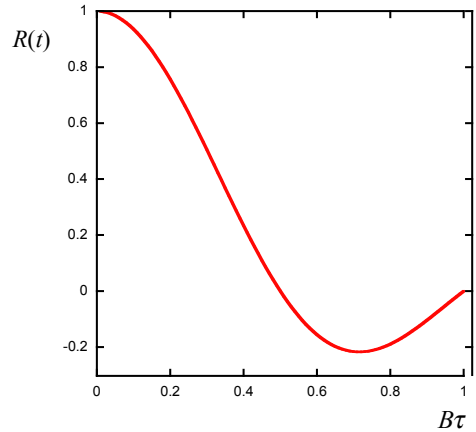

Figure 2. The autocorrelation function of white noise with spectrum $S_o$ which is bandlimited at the high-frequency end with bandwidth *B*. The shifting time is $\tau$.

## 3. Defense against arbitrary clock attacks; Robust time synchronization with integrity check

Assume a generic communication channel with fixed propagation time $\tau_0$ in both directions, see Figure 3. Suppose, Alice's absolute time is *t* and Bob's time is *t\**. Suppose Alice has the *master clock* to which Bob most synchronize. Bob's clock is off by $t_0$:

$$t^* = t + t_0. \qquad (3)$$

Below we show three possible ways of time synchronization. For educational reasons, we start with the simplest one that does not offer any defense measures against Eve.

*3.1 Undefended synchronization of absolute times*

The simplest possible synchronization signal for negligible distances is a short spike sent from Alice to Bob, while Bob terminates the cable with the wave resistance to avoid reflections. It would be suitable to synchronize oscillators at the zero-crossings of their sinusoidal outputs, and such tools are widely used in electronics. However that method alone would not be able to synchronize *absolute times*. So, the simplest scheme that we are discussing here is more involved, see Figure 3. The process is as follows:

i) At time $t_1$, Alice sends a time stamp message to Bob: "My time is $t_1$ (at the beginning of this message)".

ii) After the propagation delay $\tau$, the message arrives at Bob at his time $t_1^*$. Due to the propagation delay and Bob's time offset $t_0$, the following relation holds:

$$t_1^* = t_1 + \tau + t_0 \tag{4}$$

iii) Then Bob, at time $t_2^*$ ($> t_1^* = t_1 + \tau + t_0$) responds:

"I received your time stamp at $t_1^*$ and my time is now $t_2^*$". Note, the corresponding time at Alice is $t_2^* - t_0$ then.

iv) Alice receives this message, after the propagation delay, at time $t_2$ :

$$t_2 = t_2^* - t_0 + \tau \;. \tag{5}$$

v) Then Alice shares the $t_2$ value with Bob. They can set up the following two equations:

$$t_1^* - t_1 = +\tau + t_0 \quad \text{and} \quad t_2 - t_2^* = -t_0 + \tau \;. \tag{6}$$

The solutions of Equations 6 yield Bob's time offset and the propagation time:

$$t_0 = \frac{t_1^* - t_1 - t_2 + t_2^*}{2} \qquad (7)$$

$$\tau = \frac{t_1^* - t_1 + t_2 - t_2^*}{2} \ . \qquad (8)$$

vi) With Equations 7 and 8, Bob's clock can be corrected and the exact propagation time is determined for (partial) information about the cable integrity.

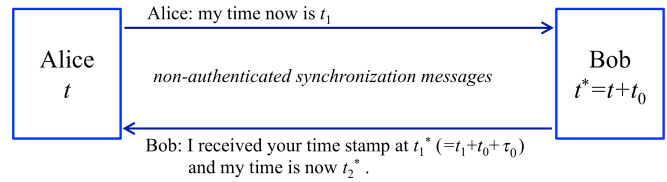

Figure 3. A simple, non-authenticated synchronization scheme for synchronizing absolute times via a classical communication channel. Alice has the master clock to which the whole system is synchronized. The absolute times of Alice and Bob are $t$ and $t^*$. The propagation delay time in the line is $\tau_0$. Bob's clock is off by $t_0$. The system allows exact synchronization of the absolute times but it is not secure. For example, Eve can remove the communicated data from the channel and she can substitute her own data to falsify the protocol.

In conclusion, the above protocol provides a satisfactory synchronization in the case of a symmetric channel with steady parameters. However, the protocol can easily be attacked. For example, Eve can remove messages and substitute her data instead and/or modify the channel and its propagation time during or between the two communications.

*3.2 Authenticated synchronization of absolute times*

To prohibit Eve to substitute her own messages into the communication protocol, the process described in Section 3.1 can be run by authenticated communications, see Figure 4. For example, Alive and Bob can digitally sign their messages by attaching a hash fingerprint of the message and encrypt that fingerprint by using a few secure bits from the last key exchange, see a more detailed description in Section 3.3. (Instead of hash, they could simply encrypt these messages but that may use up too many secure bits).

*Clock synchronization in KLJN*

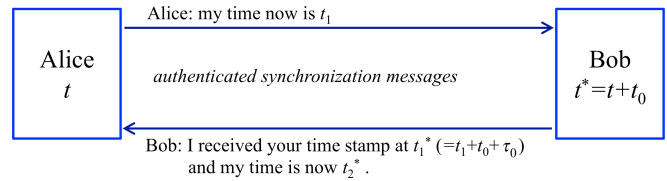

Figure 4. The authenticated version of the simple synchronization scheme in Figure 3. This system offers much better security and Eve cannot covertly substitute data; however, it is still prone to attacks, see in the text.

Thus, even though Eve can still substitute her own message, she does not know the key therefore she cannot create the encrypted hash fingerprint of that message. Thus, Alice and Bob will discover the discrepancy between the message and the encrypted hash fingerprint (either the hash of the original message or the hash of Eve's possibly substituted one). Consequently, they detect the attack.

Therefore, this method is much more robust due to the authentication. However, it is still prone to attacks: for example, Eve can change the length of the line when Bob sends his response to Alice.

*3.3 Robust and secure synchronization scheme utilizing system integrity check*

This synchronization takes place during the normal bit exchange operation of the KLJN system and it may involve a single BEP or multiple periods. The details of the process are shown below, including the steps of authentication (mentioned above), see Figures 5 and 6. The check of line and time integrity of the system is utilized for the process.

(i) As it is already required for the general defense against active attacks, during the BEP, Alice and Bob measure and store the voltage and current data indexed with their local absolute time at their terminal. They write these data into a file F($k$) where $k$ is the index number of the $k^{th}$ BEP.

(ii) They exchange F($k$) via an authenticated file transfer, that is, they also attach the encrypted version of the hash signature H($k$) of F($k$). The authentication must be unconditionally secure, which is achieved via encrypting the H($k$) hash into a $C_H(k)$ ciphertext by using a small fraction of the key bits of the former unconditionally secure key. Hashes are small strings; thus, the encryption uses up only a minor fraction of the formerly generated key. Eve cannot replace F($k$) and H($k$) because she does not know the secure key to encrypt her replacement of the hash H($k$). However, the receiving party has the shared key, thus he/she can obtain H($k$) by decrypting $C_H(k)$, and then check the authenticity of F($k$).

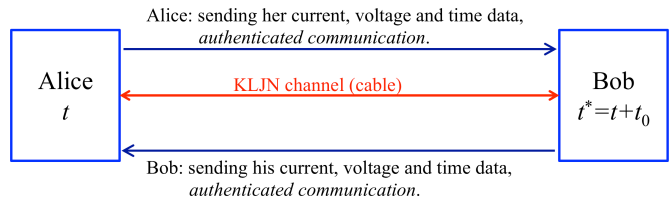

Figure 5. The robust and secure synchronization scheme integrating the classical communication channel with a KLJN line. It combines system integrity check and time synchronization of the KLJN system via the authenticated exchange of KLJN current, voltage and time data. Then Alice and Bob run simulations based on the data and selects the proper absolute time while checking the integrity of the system by comparing measured and simulated data, see also Figure 6. As an additional integrity check between subsequent simulations-based tests, short synchronization messages can also be sent by Alice and Bob, at random times, with the extra authentication protocol outlined in Figure 4.

(iii) The integrity and synchronicity of these data is checked by plugging their voltage data and the received voltage data into a cable model (simulator) and comparing the resulting current data with the measured ones. Alternatively, they can use the current data as input for the simulation and check the voltages. Both Alice and Bob are conducting these processes simultaneously. In the simplest cases (e.g., with short cable), such a cable model can be just the Ohm's law with the measured terminal voltages and the known wire resistance as input data to calculate the simulated wire current at the two ends and compare that with the shared measurement results.

The example in Figure 6 uses the measured cable voltages, $U_{cAm}(t)$ and $U_{cBm}(t^*+\Delta t^*)$, by Alice and Bob, respectively, as inputs for their cable simulators. These measured voltages and the corresponding measured currents, $I_{cAm}(t)$ and $I_{cBm}(t^*+\Delta t^*)$, of Alice and Bob, respectively, are recorded. At this stage, the time offset of Bob is unknown yet thus $t^*$ is his non-synchronized absolute time and the time shift $\Delta t^*$ is varied to test the simulated response versus the measured response. The corresponding, simulated cable currents, $I_{cAs}(t)$ and $I_{cBs}(t^*+\Delta t^*)$, of Alice and Bob are compared with the corresponding measured cable currents, $I_{cAm}(t)$ and $I_{cBm}(t^*)$, respectively. If the measured and simulated currents differ, there is either an ongoing active attack that modified the line or the times have asynchrony. The particular time shift $\Delta t^*$, where the measured and simulated currents match, gives Bob's exact time offset,

$$t_0 = -\Delta t^* , \tag{9}$$

see Equation 3. Learning the $t_0$ value allows Bob to synchronize his clock to Alice's master clock, thus they have synchronized absolute times.

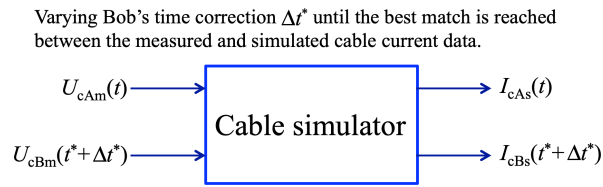

Figure 6. The simulation and comparison part of the scheme shown in Figure 5. The $\Delta t^*$ value that provides the best fit between the measured and simulated currents yields Bob's $t_0$ time offset, see Equation 9.

iv) Interestingly, the method described above does not directly provide the propagation time constant $\tau$. As an additional security measure, the simple, authenticated synchronization described in Section 3.2 can also be done at random times. That will provide the $\tau$ and $t_0$. The $t_0$ offset should be idealistically zero after the integrity-based synchronization. If not, and/or $\tau$ is different from the nominal value then there is an ongoing attack. Idealistically, to approve the integrity of the channel and the time synchronization, the KLJN scheme should pass both checks.

**Conclusion**

Efficient data comparison defense methods against active attacks in the KLJN system require that the timing of the data and operations of Alice and Bob are synchronized. We introduced a robust clock synchronization protocol (Section 3.3, (i)-(iv)), based on the measurement data that Alice and Bob are sharing for defenses against arbitrary active attacks followed by authenticated synchronization communications (Section 3.2). The agreements of the simulated and measured data, and monitoring the propagation time $\tau$, guarantee the synchronicity, system integrity and the detection of active attacks.

**Acknowledgement**

The Author is grateful to Oliver Cohen for bringing up the question of time synchronization of KLJN key exchangers, and showing some relevant papers about the same issues of QKD.

*Clock synchronization in KLJN*[53] L. B. Kish, Response to Feng Hao's paper "Kish's key exchange scheme is insecure," Fluctuation and Noise Letters, 6 (2006) C37-C41.

[54] L. B. Kish, Protection against the man-in-the-middle-attack for the Kirchhoff-loop-Johnson (-like)-noise cipher and expansion by voltage-based security, Fluctuation and Noise Letters, 6 (2006) L57-L63.

[55] L. J. Gunn, A. Allison, and D. Abbott, A new transient attack on the Kish key distribution system, IEEE Access, 3 (2015) 1640-1648.

[56] L. B. Kish and C. G. Granqvist, Comments on "a new transient attack on the Kish key distribution system", Metrology and Measurement Systems, 23 (2015) 321-331.

[57] L. J. Gunn, A. Allison, and D. Abbott, A directional wave measurement attack against the Kish key distribution system, Scientific Reports, 4 (2014) 6461.

[58] H. P. Chen, L. B. Kish, and C. G. Granqvist, On the "cracking" scheme in the paper "a directional coupler attack against the Kish key distribution system" by Gunn, Allison and Abbott, Metrology and Measurement Systems, 21 (2014) 389-400.

[59] H. P. Chen, L. B. Kish, C. G. Granqvist, and G. Schmera, Do electromagnetic waves exist in a short cable at low frequencies? What does physics say? Fluctuation and Noise Letters, 13 (2014) 1450016.

[60] L. B. Kish, Z. Gingl, R. Mingesz, G. Vadai, J. Smulko, and C. G. Granqvist, Analysis of an attenuator artifact in an experimental attack by Gunn-Allison-Abbott against the Kirchhoff-law-Johnson-noise (KLJN) secure key exchange system, Fluctuation and Noise Letters, 14 (2015) 1550011.

[61] L. B. Kish, D. Abbott, and C. G. Granqvist, Critical analysis of the Bennett-Riedel attack on secure cryptographic key distributions via the Kirchhoff-law-Johnson-noise scheme, PloS One, 8(12) (2013) e81810.

[62] C. H. Bennett and G. Brassard. Quantum cryptography: Public key distribution and coin tossing. In Proceedings of IEEE International Conference on Computers, Systems and Signal Processing, volume 175, p. 8. New York, 1984.

[63] H. P. Yuen, Security of quantum key distribution, IEEE Access, 4 (2016) 724-749.

[64] H. P. Yuen, Essential elements lacking in security proofs for quantum key distribution, Proc. SPIE, 8899 (2013) 88990J-88990J-13.

[65] H. P. Yuen, Essential lack of security proof in quantum key distribution, arXiv preprint https://arxiv.org/abs/1310.0842. (2013).

[66] O. Hirota, Incompleteness and limit of quantum key distribution theory, arXiv preprint https://arxiv.org/abs/1208.2106. (2012).

[67] S. Sajeed, A. Huang, S. Sun, F. Xu, V. Makarov, and M. Curty, Insecurity of detector-device-independent quantum key distribution, Physical Review Letters, 117(25) (2016) 250505.

[68] I. Gerhardt, Q. Liu, A. Lamas-Linares, J. Skaar, C. Kurtsiefer, and V. Makarov, Full-field implementation of a perfect eavesdropper on a quantum cryptography system, Nature Communications, 2(349) (2011) 1-6.

[69] L. Lydersen, C. Wiechers, C. Wittmann, D. Elser, J. Skaar, and V. Makarov, Hacking commercial quantum cryptography systems by tailored bright illumination, Nature Photonics, 4(10) (2010) 686-689.

[70] I. Gerhardt, Q. Liu, A. Lamas-Linares, J. Skaar, V. Scarani, V. Makarov, and C. Kurtsiefer, Experimentally faking the violation of Bell's inequalities, Physical Review Letters, 107 (2011) 170404.

[71] V. Makarov and J. Skaar, Fakes states attack using detector efficiency mismatch on SARG04, phase-time, DPSK, and Ekert Protocols, Quantum Information & Computation, 8(6) (2008) 622-635.

[72] C. Wiechers, L. Lydersen, C. Wittmann, D. Elser, J. Skaar, C. Marquardt, V. Makarov, and G. Leuchs, After-gate attack on a quantum cryptosystem, New Journal of Physics, 13 (2011) 013043.

[73] L. Lydersen, C. Wiechers, C. Wittmann, D. Elser, J. Skaar, and V. Makarov, Thermal blinding of gated detectors in quantum cryptography, Optics Express, 18(26) (2010) 27938-27954.
14